\documentclass[12pt]{article}
\newcommand{\be}{\begin{equation}}
\newcommand{\ee}{\end{equation}}
\newcommand{\bea}{\begin{eqnarray}} 
\newcommand{\eea}{\end{eqnarray}}
\newcommand{\nn}{\nonumber}
\newcommand{\lb}{\label}
\newcommand{\lcb}{\left\{}

\newcommand{\lsb}{\left[}
\newcommand{\rcb}{\right\}}

\newcommand{\rsb}{\right]}
\newcommand{\fr}{\frac}
\newcommand{\half}{\fr{1}{2}} 
\newcommand{\cC}{{\cal C}}
\newcommand{\cJ}{{\cal J}}
\newcommand{\cK}{{\cal K}}
\newcommand{\cL}{{\cal L}} 
\begin{document}
\begin{center}
{\bf JORDAN-SCHWINGER-TYPE REALIZATIONS OF 

\smallskip

THREE-DIMENSIONAL POLYNOMIAL ALGEBRAS}

\bigskip 

V. SUNIL KUMAR and B. A. BAMBAH,$^*$ \\ 

\smallskip 

{\em School of Physics, University of Hyderabad \\
Hyderabad - 500046, India} \\ 
$^*${\em bbsp@uohyd.ernet.in} \\ 

\medskip 

R. JAGANNATHAN 

\smallskip 

{\em The Institute of Mathematical Sciences \\
C.I.T. Campus, Tharamani, Chennai - 600113, India} \\  
{\em jagan@imsc.ernet.in}  
\end{center}

\bigskip

\begin{quote}
A three-dimensional polynomial algebra of order $m$ is defined by the commutation 
relations $[P_0 , P_\pm]$ $=$ $\pm P_\pm$, $[P_+ , P_-]$ $=$ $\phi^{(m)}(P_0)$ 
where $\phi^{(m)}(P_0)$ is an $m$-th order polynomial in $P_0$ with the 
coefficients being constants or central elements of the algebra.  It is shown 
that two given mutually commuting polynomial algebras of orders $l$ and $m$ 
can be combined to give two distinct $(l+m+1)$-th order polynomial algebras.  
This procedure follows from a generalization of the well known Jordan-Schwinger 
method of construction of $su(2)$ and $su(1,1)$ algebras from two mutually 
commuting boson algebras.

\medskip 


\smallskip 


\end{quote}

\bigskip

\section{Introduction}	
\renewcommand{\theequation}{1.\arabic{equation}}
\setcounter{equation}{0}
A three-dimensional polynomial algebra of order $m$ is defined by the commutation 
relations 
\bea 
[P_0 , P_\pm] & = & \pm P_\pm, \nn \\  
{}[P_+ , P_-] & = & \phi^{(m)}(P_0) 
 = \sum_{j=0}^m a_j P_0^{~j}, \quad a_m \neq 0, 
\lb{padef}
\eea 
where the coefficients $\lcb a_j \rcb$ are constants or central elements of the 
algebra.  It should be noted that in this case the Jacobi identity does not impose 
any restriction on the values of the coefficients $\lcb a_j \rcb$.  In general, 
\be
\cC = P_+P_- + g(P_0 - 1) = P_-P_+ + g(P_0), 
\ee
is a Casimir operator of the algebra (\ref{padef}) where $g(P_0)$ 
is defined by the relation 
\be
g(P_0) - g(P_0 - 1) = \phi^{(m)}(P_0). 
\lb{g}
\ee 
Note that $g(P_0)$ is an $(m+1)$-th order polynomial in $P_0$ and can be 
determined uniquely, up to an additive constant, from the relation (\ref{g}).  In 
the following we shall take the $g$-function to be defined uniquely without the 
constant term.  

It may be noted that the canonical boson algebra corresponds to the case $m$ $=$ $0$ 
when $\phi^{(0)}$ is just a constant.  The $su(2)$ and $su(1,1)$ algebras belong to 
$m$ $=$ $1$ corresponding to a monomial $\phi^{(1)}$.  If $m$ $=$ $2$ we have a 
quadratic algebra and if $m$ $=$ $3$ we have a cubic algebra.  A well known cubic 
algebra is the Higgs algebra$^1$ with the commutation relations 
\be 
[H_0 , H_\pm] = \pm H_\pm,  \qquad 
[H_+ , H_-] = 4hH_0^3 + 2aH_0, 
\lb{higgs}
\ee 
where $h$ can be positive or negative.  Such polynomial algebras, and their 
supersymmetric versions including anticommutation relations, represent the 
nonlinear symmetry or dynamical algebras in several physical problems in quantum 
mechanics, statistical physics, field theory, Yang-Mills-type gauge theories, 
integrable systems, quantum optics, etc. ({\em e.g.}, see Refs.[1-27]).  
Hence a general mathematical study of such algebras is of interest.  Here we shall 
be concerned only with polynomially deformed three dimensional Lie algebras and 
present a Jordan-Schwinger-like method of combining lower order polynomial algebras 
to get higher order polynomial algebras generalizing the earlier works on the 
Higgs algebra$^{10}$ and quadratic algebras$^{25,26}$.   

\section{$su(2)$ and $su(1,1)$} 
\renewcommand{\theequation}{2.\arabic{equation}}
\setcounter{equation}{0}
Let us briefly recall the construction of $su(2)$ and $su(1,1)$ algebras starting 
with two boson algebras.  Let $(a_+ , a_-)$ and $(b_+ , b_-)$ be two mutually  
commuting boson creation-annihilation operator pairs.  Let $N_a$ $=$ $a_+a_-$ and 
$N_b$ $=$ $b_+b_-$ be the corresponding number operators.  As is well known, 
$(J_0, J_+, J_-)$ defined by 
\be
J_0 = \half(N_a - N_b), \qquad 
J_+ = a_+ b_-, \qquad
J_- = a_- b_+, 
\ee 
satisfy the $su(2)$ algebra, 
\be
\lsb J_0 , J_\pm \rsb = \pm J_\pm, \qquad
\lsb J_+ , J_- \rsb = 2J_0.  
\ee
In this Jordan-Schwinger realization of $su(2)$, $N_a + N_b$ is seen to be a 
central element\,: if 
\be
\cL_J = \half(N_a + N_b), 
\ee
then,
\be
\lsb \cL_J , J_{0,\pm} \rsb = 0.  
\ee
The $g$-function in this case is $g(J_0)$ $=$ $J_0(J_0+1)$ and hence the Casimir 
operator is  
\be 
\cC_J = J_+J_- + J_0(J_0 - 1) = J_-J_+ + J_0(J_0+1). 
\ee

In an analogous way, $(K_0, K_+, K_-)$ defined by  
\be
K_0 = \half(N_a + N_b), \qquad
K_+ = a_+b_+, \qquad
K_- = a_-b_-,  
\ee
satisfy the algebra  
\be
\lsb K_0 , K_\pm \rsb = \pm K_\pm, \qquad 
\lsb K_+ , K_- \rsb = -(2K_0 + 1).
\ee 
Calling $K_0+\half$ as $K_0$ this algebra becomes the standard $su(1,1)$ algebra 
\be
\lsb K_0 , K_\pm \rsb = \pm K_\pm, \qquad 
\lsb K_+ , K_- \rsb = -2K_0. 
\ee 
Now, 
\be
\cL_K = \half(N_a - N_b) 
\ee
is a central element of the algebra\,:
\be
\lsb \cL_K , K_{0,\pm} \rsb = 0.  
\ee 
The $g$-function is $g(K_0)$ $=$ $-K_0(K_0+1)$ and hence the Casimir operator is 
\be 
\cC_K = K_+ K_- - K_0(K_0-1) = K_-K_+ - K_0(K_0+1). 
\ee 

\section{Jordan-Schwinger-like construction of polynomial algebras} 
\renewcommand{\theequation}{3.\arabic{equation}}
\setcounter{equation}{0}
Let us now generalize the above construction of $su(2)$ and $su(1,1)$ algebras 
leading to polynomial algebras.  This work follows from an observation in Ref.[10] 
on the construction of the Higgs cubic algebra (\ref{higgs}) starting with mutually 
commuting $su(2)$ and $su(1,1)$ algebras (when $h$ $>$ $0$) or two mutually commuting 
$su(1,1)$ algebras (when $h$ $<$ $0$) and also a generalization of our earlier 
work$^{25,26}$ in which we have constructed four classes of quadratic algebras 
combining a boson algebra with $su(2)$ and $su(1,1)$.  Let $(L_0 , L_\pm)$ and 
$(M_0 , M_\pm)$ be the generating sets of two mutually commuting polynomial algebras 
of order $l$ and $m$, respectively.  Then, using these two algebras as building 
blocks, we can construct two distinct polynomial algebras of order $l+m+1$ 
analogous to $su(1,1)$ and $su(2)$.  To this end we proceed as follows.  

Let 
\be 
\cJ_0 = \half(L_0 - M_0), \qquad 
\cJ_+ = \mu L_+M_-, \qquad 
\cJ_- = \mu L_-M_+, 
\ee  
in analogy with with $su(2)$.  Then, it is easily seen that these generate an 
algebra with the commutation relations :  
\bea
[\cJ_0 , \cJ_\pm] & = & \pm\cJ_\pm, \nn \\
{}[\cJ_+ , \cJ_-] & = & 
\mu^2\{[\cC_M-g_M(\cL_{\cJ}-\cJ_0-1)]\phi^{(l)}(\cL_{\cJ}+\cJ_0) \nn\\ 
                  &   & \ \  - [\cC_L-g_L(\cL_{\cJ}+\cJ_0-1)]\phi^{(m)}
                              (\cL_{\cJ}-\cJ_0)\}, 
\lb{jalg}
\eea 
where 
\be
\cL_{\cJ} = \half(L_0 + M_0) 
\ee 
is a central element of the algebra and $\cC_L$, $\cC_M$, $g_L$, and $g_M$ are the 
Casimir operators and the $g$-functions of the $L$-algebra and the $M$-algebra, 
respectively.  Let us call this algebra (\ref{jalg}) as $\cJ$.  It is 
straightforward to see that $\cJ$ is a polynomial algebra of order $l+m+1$.   
 
Now, in analogy with $su(1,1)$, let 
\bea 
\cK_0 = \half(L_0 + M_0), \qquad 
\cK_+ = \mu L_+M_+, \qquad 
\cK_- = \mu L_-M_-. 
\eea 
The corresponding commutation relations are : 
\bea
[\cK_0 , \cK_\pm] & = & \pm\cK_\pm, \nn \\
{}[\cK_+ , \cK_-] & = & \mu^2 \{[\cC_L - g_L(\cK_0+\cL_{\cK}-1)] 
                        \phi^{(m)}(\cK_0-\cL_{\cK}) \nn \\
                  &   & \quad + [\cC_M - g_M(\cK_0-\cL_{\cK})] 
                        \phi^{(l)}(\cK_0+\cL_{\cK})\},  
\lb{kalg}
\eea 
where 
\be
\cL_{\cK} = \half(L_0 - M_0) 
\ee 
is a central element of the algebra and $\cC_L$, $\cC_M$, $g_L$ and $g_M$ are the 
Casimir operators and the $g$-functions of the $L$-algebra and the $M$-algebra, 
respectively.  Let us call the algebra (\ref{kalg}) as $\cK$.  It is clear that 
the polynomial algebra $\cK$, distinct from $\cJ$, is also of order $l+m+1$.   

\section{Examples}
\renewcommand{\theequation}{4.\arabic{equation}}
\setcounter{equation}{0}
Let us first consider an example from our earlier work$^{25,26}$.  With 
$(J_0, J_\pm)$ as the generators of the $su(2)$ algebra and $(a_+,a_-,N)$ as the 
boson operators it can be easily verified that 
\be
Q_0 = \half(J_0 - N), \qquad 
Q_+ = \mu J_+a_-, \qquad 
Q_- = \mu J_-a_+   
\ee
satisfy a quadratic algebra 
\bea 
[Q_0 , Q_\pm] & = & \pm Q_\pm, \nn \\
{}[Q_+ , Q_-] & = & -\mu^2\{3Q_0^2 + (2\cL_Q - 1)Q_0 - [\cC_J + \cL_Q(\cL_Q + 1)]\},  
\eea 
where 
\be
\cL_Q = \half(J_0 + N)
\ee
is a central element of the algebra and $\cC_J$ is the $su(2)$ Casimir operator.  
This is a $\cJ$-type quadratic algebra resulting from the fusion of the $su(2)$ 
algebra and a boson algebra.  Fusion of a boson algebra ($l$ $=$ $0$) with $su(2)$ 
and $su(1,1)$ algebras ($m$ $=$ $1$) in this way leads to four classes of quadratic 
algebras which have been studied in detail by us earlier$^{25,26}$. 

Let us now consider a few other examples.  First, let us start with a quadratic 
algebra ($l$ $=$ $2$) and combine it with a boson algebra ($m$ $=$ $0$) to get a 
cubic algebra ($l+m+1$ $=$ $3$).  Any quadratic algebra is of the generic form 
\be 
[Q_0 , Q_\pm] = \pm Q_\pm, \qquad 
[Q_+ , Q_-] = aQ_0^2 + bQ_0 + c,  
\ee 
where $(a,b,c)$ commute with $(Q_0 , Q_\pm)$.  The corresponding $g$-function is 
\be
g(Q_0) = \fr{a}{3}Q_0^3 + \half(a+b)Q_0^2 + \fr{1}{6}(a+3b+6c)Q_0, 
\ee
and the Casimir operator is 
\bea
\cC_Q & = & Q_+Q_- + \fr{a}{3}Q_0^3 - \half(a-b)Q_0^2 
                   + \fr{1}{6}(a-3b+6c)Q_0 - \fr{1}{3}(a-3c-1)  \nn \\
      & = & Q_-Q_+ + \fr{a}{3}Q_0^3 + \half(a+b)Q_0^2 + \fr{1}{6}(a+3b+6c)Q_0. 
\eea  
Following the procedure prescribed above we define 
\be 
C_0 = \half(Q_0 - N), \qquad 
C_+ = \mu Q_+a_-, \qquad 
C_- = \mu Q_-a_+ . 
\ee 
and
\be
\cL_C = \half(Q_0 + N). 
\ee 
Then, we get the cubic algebra 
\bea
[C_0 , C_\pm] & = & \pm C_\pm, \nn \\
{}[C_+ , C_-] & = & -\mu^2\lcb \fr{4a}{3}C_0^3 + \half(4a\cL_C-a+3b)C_0^2 \right.
                     \nn \\
              &   & \ \ - \lsb (a-b)\cL_C - \fr{1}{6}(a-3b+12c)\rsb C_0 \nn \\
              &   & \ \ - \lsb C_Q + a\cL_C^3 + \half(a+b)\cL_C^2 - 
                          \fr{1}{6}(a-3b)\cL_C \right. \nn \\ 
              &   & \ \ \left.\left. -\fr{1}{3}(a-3c-1)\rsb\rcb, 
\eea
where $\cL_C$ is a central element of the algebra.  

As the next example, let us combine mutually commuting $su(2)$ and $su(1,1)$ 
algebras to get the Higgs cubic algebra (\ref{higgs}) with $h$ $>$ $0$, following 
Ref.[10].  Let $(J_0,J_\pm)$ and $(K_0,K_\pm)$ be, respectively, the generators of 
mutually commuting $su(2)$ and $su(1,1)$ algebras.  Then,   
\be 
H_0 = \half(J_0 - K_0), \qquad 
H_+ = \mu J_+K_-, \qquad 
H_- = \mu J_-K_+, 
\ee 
generate the cubic algebra 
\bea
[H_0 , H_\pm] & = & \pm H_\pm, \nn \\
{}[H_+ , H_-] & = & \mu^2 \lcb 4H_0^3 +  
                     \lsb 2(\cC_J-\cC_K) + 4\cL_H^2 \rsb H_0 + 2(\cC_J 
                     + \cC_K)\cL_H\rcb, \nn \\
  &  & 
\eea
where 
\be
\cL_H = \half(J_0 + K_0)
\ee
is a central element of the algebra and $\cC_J$ and $\cC_K$ are the Casimir 
operators of the $J$ and $K$ algebras, respectively.  The Higgs algebra 
(\ref{higgs}) with $h$ $>$ $0$ can be now identified with this cubic algebra by 
taking $\mu^2$ $=$ $h$ and $\cC_J+\cC_K$ $=$ $0$, and suitably choosing the value 
of $\cL_H$.  

As the final example, let us combine two mutually commuting $su(1,1)$ algebras to 
get the Higgs cubic algebra (\ref{higgs}) with $h$ $<$ $0$, again following Ref.[10].  
Let $(L_0,L_\pm)$ and $(M_0,M_\pm)$ be the generators of two mutually commuting 
$su(1,1)$ algebras.  Then,  
\be
H_0 = \half(L_0 - M_0), \qquad 
H_+ = \mu L_+M_-, \qquad 
H_- = \mu L_-M_+, 
\ee 
generate the cubic algebra 
\bea
[H_0 , H_\pm] & = & \pm H_\pm, \nn \\
{}[H_+ , H_-] & = & -\mu^2 \lcb 4H_0^3 +  
                     \lsb 2(\cC_L+\cC_M) - 4\cL_H^2 \rsb H_0 
                     - 2(\cC_L-\cC_M)\cL_H\rcb, \nn \\
  &  & 
\eea
where 
\be
\cL_H = \half(L_0 + M_0)
\ee
is a central element of the algebra $\cC_L$ and $\cC_M$ are the Casimir operators 
of the $L$ and $M$ algebras respectively.  The Higgs algebra (\ref{higgs}) with 
$h$ $<$ $0$ can be now identified with this cubic algebra by taking $\mu^2$ $=$ 
$|h|$ and $\cC_L$ $=$ $\cC_M$, and suitably choosing the value of $\cL_H$.  

The last two examples show that the observation in Ref.[10] that the Higgs algebra 
can be obtained by combining mutually commuting $su(2)$ and $su(1,1)$ algebras, or 
two mutually commuting $su(1,1)$ algebras, in the Jordan-Schwinger way, is a 
special case of a generalized Jordan-Schwinger method of constructing polynomial 
algebras.  Now, it should be noted that two distinct cubic algebras are obtained 
whenever a boson algebra is combined with a quadratic algebra or an $su(2)$ or 
$su(1,1)$ algebra is combined with another $su(2)$ or $su(1,1)$ algebra.  Thus it 
is clear that there are several classes of cubic algebras of which the Higgs  
algebra is a special case.  It should be interesting to study these algebras in 
detail. 
 
\section{Conclusion}
Generalizing the method of construction of the Higgs algebra found in Ref.[10] and 
the method of construction of quadratic algebras described in Refs.[25,26] we have 
shown how two mutually commuting polynomial algebras of order $l$ and $m$ can be 
combined in the Jordan-Schwinger way to get two distinct polynomial algebras of 
order $l+m+1$.  The simplest example of this construction is the Jordan-Schwinger 
realization of $su(2)$ and $su(1,1)$, linear algebras corresponding to order $m$ 
$=$ $1$, starting with two commuting boson algebras which are algebras of order 
$m$ $=$ $0$.  By combining a boson algebra with $su(2)$ or $su(1,1)$ we get four 
classes of quadratic algebras.  Combining a boson algebra and these quadratic 
algebras or combining an $su(2)$ or $su(1,1)$ algebra with another $su(2)$ or 
$su(1,1)$ algebra one can get several classes of cubic algebras of which the 
Higgs algebra is a special case.  Higher order algebras can be generated 
similarly by combining lower order algebras.  It should be noted that the above 
construction leads to polynomial algebras in which the coefficients of the 
polynomials are central elements which are defined in terms of the Casimir 
operators of the original algebras with which one starts or a combination of 
their generators $L_0$ and $M_0$.  This construction also helps find some 
irreducible representations of the constructed three dimensional polynomial 
algebras starting with the irreducible representations of the underlying $L$ and 
$M$ algebras.  For example, in the case of each of the four classes of quadratic 
algebras we have obtained in Refs.[25,26] some irreducible representations have 
been found starting with the irreducible representations of the $su(2)$, 
$su(1,1)$ and boson algebras.  Then, an interesting problem, which would help 
understand the classification and representation theory of three dimensional 
polynomial algebras, is : Given a three dimensional polynomial algebra with 
certain numerical coefficients is it possible to identify it with a particular 
type of three dimensional polynomial algebra generated by the fusion of two 
lower order algebras and corresponding to certain numerical values of the central 
elements?


\bigskip 

\begin{thebibliography}{99}
\bibitem{H}
P. W. Higgs, {\it J. Phys. A: Math. Gen.} {\bf 12} 309 (1979).
\bibitem{L}
M. Lakshmanan and K. Eswaran, {\it J. Phys. A: Math. Gen.} {\bf 8} 1658 (1975). 
\bibitem{S}
E. K. Sklyanin, {\it Funct. Anal. Appl.} {\bf 16} 263 (1982). 
\bibitem{C} 
T. Curtwright and C. Zachos, {\it Phys. Lett. B} {\bf 243} 237 (1990).    
\bibitem{P}
A. P. Polychronakos, {\it Mod. Phys. Lett. A} {\bf 5} 2325 (1990).  
\bibitem{R}
M. Ro\v{c}ek, {\it Phys. Lett. B} {\bf 255} 554 (1991).  
\bibitem{Ga} 
O. F. Gal'bert, Ya. I. Granovskii and A. S. Zhedanov, {\it Phys. Lett. A} {\bf 153} 
177 (1991).  
\bibitem{G} 
Ya. I. Granovskii, A. S. Zhedanov and I. M. Lutzenko, {\it J. Phys. A: Math. Gen.} 
{\bf 24} 3887 (1991).  
\bibitem{Sh} 
K. Schoutens, A. Sevrin and P. van Nieuwenhuizen, {\it Phys. Lett. B} {\bf 255} 
549 (1991).   
\bibitem{Z}
A. S. Zhedanov, {\it Mod. Phys. Lett. A} {\bf 7} 507 (1992).   
\bibitem{K}
V. P. Karassiov, {\it J. Sov. Laser Res.} {\bf 13} 188 (1992).   
\bibitem{B} 
D. Bonatsos, C. Daskaloyannis and K. Kokkotas K, {\it Phys. Rev. A} {\bf 48} 
3407 (1993).   
\bibitem{Q} 
C. Quesne, {\it Phys. Lett. A} {\bf 193} 245 (1994).  
\bibitem{Le} 
P. L\'{e}tourneau and L. Vinet, {\it Ann. Phys.} {\bf 243} 144 (1995).  
\bibitem{Jv}
J. Van der Jeugt and R. Jagannathan, {\it J. Math. Phys.} {\bf 36} 4507 (1995).  
\bibitem{A} 
B. Abdesselam, J. Beckers, A. Chakrabarti and N. Debergh, {\it J. Phys. A:  Math. 
Gen.} {\bf 29} 3075 (1996).   
\bibitem{Bo} 
J. de Boer, F. Harmsze and T. Tijn, {\it Phys. Rep.} {\bf 272} 139 (1996).   
\bibitem{M}
V. I. Man'ko, G. Marmo, E. C. G. Sudarshan and F. Zaccaria, {\it Phys. Scr.} {\bf 55} 
528 (1997).   
\bibitem{F} 
D. J. Fern\'{a}ndez and V. Hussin, {\it J. Phys. A: Math. Gen.} {\bf 32} 3603 
(1999).    
\bibitem{Ro}
B. Roy and P. Roy, {\it Quantum Semiclass. Opt.} {\bf 1} 341 (1999).   
\bibitem{Q2}
C. Quesne, {\it Phys. Lett. A} {\bf 272} 313 (2000). ({\it Erratum}: {\bf 275} 
313 (2000)).  
\bibitem{SBJPS}
V. Sunil Kumar, B. A. Bambah, R. Jagannathan, P. K. Panigrahi and V. Srinivasan,   
{\it Quantum Semiclass. Opt.} {\bf 1} 126 (2000).  
\bibitem{KP}
S. M. Klishevich and M. S. Plyushchay, {\it Nucl. Phys. B} {\bf 616} 403 (2001). 
\bibitem{DYSK}
J. Delgado, E. C. Yustas, L. L. S\'{a}nchez-Soto and A. B. Klimov, 
{\tt arXiv:quant-ph/0102026}.  
\bibitem{SBJ}
V. Sunil Kumar, B. A. Bambah and R. Jagannathan, {\it J. Phys. A: Math. Gen.} {\bf 34} 
8583 (2001).  
\bibitem{Su} 
V. Sunil Kumar, {\em Aspects of polynomial algebras and their physical applications}, 
Ph.D. Thesis (submitted to University of Hyderabad, Hyderabad, India, 2002), 
{\tt arXiv:math-ph/0203047}  
\bibitem{JM}
L. Jonke and S. Meljanac, {\tt arXiv:hep-th/0203245}.  
\end{thebibliography}
\end{document}